\documentclass[prl,aps,amssymb,amsmath,superscriptaddress,twocolumn,showpacs,
letter]{revtex4}

\usepackage{graphicx}

\renewcommand{\epsilon}{\varepsilon}

\renewcommand{\phi}{\varphi}

\begin{document}

\title{Possible Lattice Distortions in the Hubbard Model for Graphene}
\thanks{\copyright\, 2011 by the authors. This paper may be
reproduced, in its entirety, for non-commercial purposes.}

\author{Rupert L. Frank}
\email{rlfrank@math.princeton.edu}
\affiliation{Department of Mathematics,
Princeton University, Washington Road, Princeton, NJ 08544, USA}

\author{Elliott H. Lieb}
\email{lieb@princeton.edu}
\affiliation{Department of Mathematics,
Princeton University, Washington Road, Princeton, NJ 08544, USA}
\affiliation{Department of Physics,
Princeton University, P.~O.~Box 708, Princeton, NJ 08542, USA}

\begin{abstract}
  The Hubbard model  on the honeycomb lattice is a well
 known model for  graphene. Equally well known is the
Peierls type of instability of the lattice bond lengths. 
In the context of these two approximations we ask and
answer the question of the possible  lattice distortions
for graphene in zero magnetic field. The answer is that  
in the thermodynamic limit only periodic, reflection-symmetric
distortions are
allowed and these have at most 6 atoms per unit cell as
compared to two atoms for the undistorted lattice.
\end{abstract}


\maketitle


Graphene is a two-dimensional array of carbon atoms arranged in a
honeycomb lattice. The number of delocalized electrons, $N$, equals the number
of carbon atoms, i.e., the conduction band is half-filled. The dynamics of the
electrons is often modelled by the Hubbard Hamiltonian
\begin{align}
\label{eq:ham}
H(T)= & - \sum_{\sigma=\uparrow,\downarrow} \sum_{ \langle x,y\rangle}
t_{xy} \left( c^\dagger_{x,\sigma} c_{y,\sigma} + h.c. \right) \\
& + U \sum_{x} \left(n_{x \uparrow} -\frac12\right) \left(n_{x
\downarrow} -\frac12\right) + \sum_{ \langle x,y\rangle} F(t_{xy}) \,. \nonumber
\end{align}
The $c_{x,\sigma}$ are fermion annihilation operators and
$n_{x,\sigma}=c^\dagger_{x,\sigma}c_{x,\sigma}$. For the purposes of this paper
the on-site repulsion $U$ can have any sign as
long as it is the same for all lattice sites. In particular, for $U=0$ we have
the H\"uckel model.

The notation emphasizes the
dependence on the near\-est-neighbor hopping matrix $T=(t_{xy})$. We
take the hopping matrix elements $t_{xy}$ to be positive, although not
necessarily independent of the pair $x,y$.
(By the well-known particle-hole transformation on a bipartite lattice, we
could, as well, take the $t_{xy}$'s negative.) It is important for us that
$t_{xy}$ is real, i.e., that there is no magnetic field.

The interesting quantity, which does not usually appear in the Hubbard model, is
the distortion energy $F(t_{xy})$. What we are assuming is that $t_{xy}$
depends on the physical distance $d_{xy}$ between the lattice sites $x$ and
$y$. There is an equilibrium value $d^{(0)}$ and deviations from this value
cost a positive energy. After eliminating $d_{xy}$ we can assume that the
distortion energy depends on $t_{xy}$. This is the quantity $F(t_{xy})$ in
\eqref{eq:ham}. Thus, $F(t)$ has a minimum at $t=t^{(0)}$, corresponding to
$d^{(0)}$, and it goes to infinity as $t$ goes to zero (infinitely separated
atoms) or to infinity (no atomic separation). We do not have to assume that $F$
is convex. This model, while quite general, does not take account of the
possible distortion energy
connected with the change in bond angles or possible non-planar atomic
configurations, i.e., curvature of the lattice. It also does not take account
of longer range interactions \cite{HoChMu,GiMaPo}.

The problem we address is the possible configurations $T=(t_{xy})$ that
minimize the ground state energy $E(T)$ of $H(T)$. We answer this in the
thermodynamic limit by proving that one ends up with a relatively simple
periodic configuration. Nothing more complicated can occur! This periodic
configuration is shown in
Fig.~1(a), where it will be seen that the resulting lattice can have at most
three
different lengths and the unit cell can have at most six atoms. A special case
is the so-called Kekul\'e lattice in which two of the lengths are equal. It also
has
six atoms per unit cell. Many authors have considered various possible
energy-minimizing periodic distortions. The discussion can and should also be
extended to include non-periodic and chaotic structures. Our contribution is to
rigorously exclude
anything more complicated than that shown in
Fig.~1(a), in the thermodynamic limit. In particular, the ALT structure of Ref.
\cite{Mi,Ok,Ch} in Fig. 1(b) is not an
energy-minimizer for the Hubbard model, since it breaks two of the three
reflection symmetries shown in Fig.~1(a).

\begin{figure}
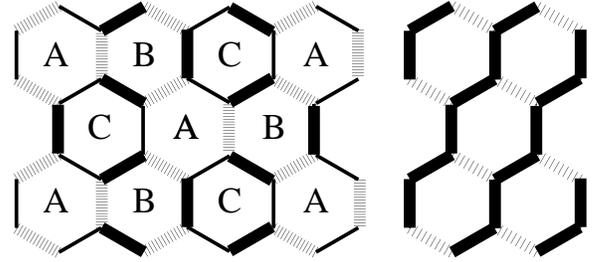

 \input{graphfig1.pstex_t}
\quad
 \input{graphfigaltb.pstex_t}
\caption{(a) On the left is the only graphene structure allowed by our theorem.
The locations of the three possible bond lengths are indicated by heavy-solid,
light-solid and dashed lines. There are six atoms (or three hexagons) per unit
cell. The Kekul\'e structure is the special case of equal length solid
lines. (b) The ALT structure on the right breaks some of
the reflection symmetry and is not allowed by our theorem.}
\end{figure}

The idea that distortion of a lattice can lower the energy is sometimes called
the Peierls instability, but the idea is older (see \cite{LiNa}). Peierls'
essential contribution \cite{Pe} was to
realize that this distortion always occurs in 1D for an infinitely long chain
or ring of atoms. It does not always occur in higher dimensions, or in a finite
1D ring, e.g., benzene.

Let us recall the situation in one dimension, i.e., the annulenes like
benzene or polyacetylene. It has been proved \cite{KeLi,LiNa} that for a closed
ring of $N$ atoms, with Hamiltonian as in \eqref{eq:ham}, the minimum energy
configuration
is precisely the one in which there are two alternating bond lengths $l_1$ and
$l_2$, provided $N=2$ mod $4$ ($N=6,10,14,\ldots$). It is possible that
$l_1=l_2$, as is the case in benzene, but nothing more complicated than
dimerization can ever occur in this model, for any $F$-function. When $N=0$ mod
$4$ more complicated things can occur when $F$ is not a quadratic, e.g.,
for $N=4$ the minimum configuration can be a trapezoid \cite{KeLi}.
Nevertheless, as $N$ goes to infinity the dimerized energy is always
asymptotically exact. The general case is analyzed in \cite{LiNa} in detail and
the stated conclusion is even shown to apply to the so-called `Spin-Peierls
model'.

The method of proof for 1D does not generally extend to higher dimensions. It
does not extend to the square lattice unless a physically unreachable magnetic
field is imposed \cite{Li}, but, as we discover here,
it \emph{does} extend to the hexagonal lattice with zero magnetic field.
The reason for this difference, ultimately, is that the energy
minimizing magnetic flux for the square lattice is non-zero, whereas for the
hexagonal lattice it is zero, as assumed in $H(T)$ above \cite{Li}. 

Naturally, we cannot expect perfect periodicity in a finite lattice, because
there will always be edge effects. This remark motivates the following theorem,
whose proof we will outline here.

{\bf Theorem}.--- \emph{For any triplet of hopping matrix elements
$(t^{(1)},t^{(2)},t^{(3)})$ let $e(t^{(1)},t^{(2)},t^{(3)})$ denote the ground
state energy per atom in the thermodynamic limit for the Hubbard model
\eqref{eq:ham} with these $t$'s arranged as in Fig.~1(a). Let $e$ be the
minumum of $e(t^{(1)},t^{(2)},t^{(3)})$
with respect to these $t$'s. Then the ground state energy $E(T)$ of
\eqref{eq:ham} with arbitrary $t$'s on a large, finite lattice $\Lambda$, with
$|\Lambda|$ sites, satisfies
\begin{equation}
 \label{eq:main}
E(T) \geq |\Lambda| e + o(|\Lambda|) \,,
\end{equation}
where $o(|\Lambda|) \leq \mathrm{const}\, |\Lambda|^{1/2} \ln|\Lambda|$. The
same conclusion holds for the free energy at finite temperature.}

Strictly speaking, we prove that there is a minimizer (to
leading order) with the structure of Fig.~1(a), but we do not
rigorously prove uniqueness of this structure. Nevertheless
the uniqueness is evident for two reasons: One is that our
proof proceeds by showing that repeated reflection of a single line
cannot raise the energy, but in reality it surely
lowers it. The second is that the structure made this way out of
repeated reflections still has the hopping matrix elements
that were in the given row before it was reflected. At this point we
could lower the energy still further by reevaluating the
optimum $t$-values. These will surely change, and the energy
will be lower because the equation for the optimum
values, namely, $F'(t_{xy})-2\langle c_x^\dagger c_y \rangle = dE/dt_{xy}=0$
will likely no longer be satisfied with the old $t$ values. For example, the ALT
structure in Fig.~1(b)
morphs into the Fig.~1(a) structure of the Kekul\'e
type. If the $t$ value on the heavy bond did not
change, then the expectation value of $c^\dagger_x c_y$
along the distinguished bond would have to be the same as
for the undistinguished bond -- which it is not.

We recall two facts about the Hubbard model \eqref{eq:ham}, which hold even for
arbitrary $T$
on any finite, bipartite lattice $\Lambda$ with an even number of sites.

(1) Among the absolute ground states there is one with $N=|\Lambda|$=number of
sites. If $U\neq 0$, this state is unique \cite{LiNa}. This means that
we do not have to worry about constraining the particle number to be equal to
the number of lattice sites, $|\Lambda|$. This constraint is automatic.

(2) Among the ground states there is one with total spin zero \cite{Li0}.

The main tool we are going to use is \emph{reflection positivity} which
originated in quantum field theory \cite{OS}. It was later used in statistical
mechanics \cite{FSS,FL,DLS,FILS,FILSII} to
prove long range order of systems with continuous symmetry. It was also
important for static situations like the solution of the flux phase problem
\cite{Li}. It is in the latter sense that we apply it here. Our theorem above can
be rephrased by saying that the lowest energy occurs for a lattice configuration
that preserves the three-fold reflection symmetry of the hexagonal lattice, as
shown in Fig.~1(a).

Let $\Lambda$ be piece of a hexagonal lattice that is reflection symmetric
about a line $l$ that passes perpendicularly through bonds. The two halves of
the lattice will
be called $\Lambda_L$ and $\Lambda_R$. Let $T_L$ and $T_R$ denote the hopping
elements on the left and on the right side, respectively, and let $T_M$ be
those corresponding to bonds cut by $l$. Let $T_L^*$ be the reflection of $T_L$
to the right and, similarly, let $T_R^*$ be the reflection of $T_R$ to the left.
Then, symbolically, $T=(T_L,T_M,T_R)$ and we want to consider
$T^L=(T_L,T_M,T_L^*)$ and $T^R=(T_R^*,T_M,T_R)$. The corresponding ground state
energies are related by the basic inequality, whose proof is in
\cite{Li,LiNa,MaNa};
\begin{equation}
 \label{eq:rp}
E(T) \geq \tfrac12 E(T^L)+ \tfrac12 E(T^R) \,.
\end{equation}

Our way forward is now obvious. Think of an infinite hexagonal lattice. There
are three directions in space for which we can draw lines in these directions
that cut lattice bonds perpendicularly (but do not pass through sites). For
each of these three directions there
are infinitely many parallel lines with these property. We would like to use
reflection positivity for each of them to say that if the hopping matrix
elements are not as shown in Fig.~1(a), then we can lower the energy by
reflecting one half of the lattice onto
the other half and keeping the better of the two choices according to
inequality \eqref{eq:rp}. In short, if the optimum configuration did not have
the symmetry of Fig.~1(a) the energy could be lowered, thereby giving a
contradiction.

The physics is now clear. The mathematical problem is to make rigorous
sense of the lattice reflections by considering finite lattices and passing
to the thermodynamic limit in an appropriate way so that at each stage the
error we make is of lower order than the area $|\Lambda|$. This is accomplished
by imposing periodic boundary conditions on the finite lattice, but it is much
more complicated, geometrically, than a similar problem for the square
lattice (with an unrealistic magnetic field flux). There, one can impose
periodic boundary conditions in \emph{two} orthogonal directions
simultaneously. This is impossible for the hexagonal lattice since the three
reflection directions are not orthogonal to each other, and we can impose
periodic boundary conditions in only \emph{one} direction at a time; see
\cite{FILSII} for a discussion of this.

Nevertheless the problem can be solved and we sketch the solution
here for the ground state energy. The extension to finite
temperature follows by similar arguments using the reflection positivity
established in \cite{DLS,FL,FILS,Li}.

{\bf Proof sketch of the Theorem}.--- 
\emph{Step 0.} The conclusion shown in Fig.~1(a) can be
deduced from a statement about individual bonds. Any bond,
$b$, in our lattice (except for those at the boundary) has four other bonds
connected to it, as shown in Fig.~2(b).
Two of these, together with $b$, form part of a hexagon and
the other two, together with $b$, form part of a
neighboring hexagon. We say that $b$ is \emph{happy} if the
first two bonds have equal $t$-values and also the second two
have equal $t$-values (possibly different values). The
$t$-value of $b$ itself is irrelevant for this definition.
(A similar definition applies to bonds at the edge, which
have fewer than four bonds attached to them.)

Our goal is to show that every bond in a minimizing
configuration is happy, for this will imply the theorem. To
be precise we shall show that all except $|\Lambda|^{1/2} \ln |\Lambda|$ of the
bonds are happy, but this is good enough for the
thermodynamic limit.

\emph{Step 1.} We
consider an increasing sequence of $\Lambda$'s each of which
is a rectangular piece of the hexagonal lattice, with even 
side lengths of order $L$. Approximately one third of the bonds are
oriented in the $y$-direction. We periodize $\Lambda$ in the $y$-direction by
connecting the sites on the top row to the corresponding sites in the bottom
row to form a new row of hexagons, and call the new lattice $\tilde\Lambda$.
This
leads to a cylinder whose axis is parallel to the $x$-axis. The new
hopping matrix $\tilde T$ consists of the old hopping matrix $T$ and some new
positive elements $t_{xy}$ connecting the top to the bottom. The new Hubbard
Hamiltonian $\tilde H(\tilde T)$ is defined as in \eqref{eq:ham} except that
the (positive) $t_{xy}$'s which connect top to bottom are
inserted with a
\emph{minus} sign. With this convention, every hexagon on the cylinder has flux
zero and any circuit around the cylinder that encircles the cylinder axis has
length $0$ mod $4$ and has flux $\pi$ \cite{LiLo}. (The flux of a circuit is
the argument of the product of the $t$'s along this circuit.) The importance of
these
fluxes lies in the following fact: Reflection positivity requires dividing the
lattice into two pieces and this requires cutting bonds, which means cutting
closed circuits. In order to have reflection positivity every closed circuit
that is cut must have flux $\pi$ if the length of the circuit is $0$ mod $4$
and flux $0$ if it is $2$ mod $4$. This is shown in \cite{Li,MaNa}. The
length of a hexagon circuit is $6$ and the length of a circuit going around
the cylinder is always $0$ mod $4$.

The insertion of additional bonds changes the
energy from $E(T)$ to $\tilde E(\tilde T)$, the ground state energy of $\tilde
H(\tilde T)$, by at most $O(L)$, i.e., order $1/L$ per atom.

We think of this cylinder as composed of rows, parallel
to the axis, as shown in Fig.~2(a), together with their
reflections, which we shall call $\theta$-rows. Thus
$\tilde\Lambda$ is a sequence of rows interspersed with $\theta$
rows. Each row and $\theta$ row has its prescribed hopping
matrix elements. Let us denote the rows as we go around
the cylinder by $A_1$, $\theta A_2$, $A_3$, $\theta A_4$,
$\ldots$, $A_{M-1}$, $\theta A_M$ (with $M$ of order $L$).
From these we can create $M$ new configurations $\tilde
T_1$, $\ldots$, $\tilde T_M$ of matrix elements of which the
$j$-th one has the rows $A_j$, $\theta A_j$, $A_j$, $\theta
A_j$, $\ldots$. It is a well known consequence of reflection
positivity \eqref{eq:rp} (called the \emph{chessboard estimate}
\cite{OS,FL,FILS,LiNa}) that
\begin{equation}\label{eq:cb}
 \tilde E(\tilde T) \geq \frac{1}{M} \sum_{j=1}^M \tilde
E(\tilde T_j) \,,
\end{equation}
from which we see that one of the $\tilde T_j$ configurations is
energetically at least as good as the original $\tilde T$ configuration. In a
$\tilde T_j$ configuration every vertical bond is happy.

\begin{figure}
 \input{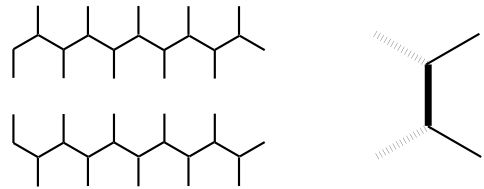}
\caption{Some definitions: (a) The top left figure is a single row and the
bottom left figure is a reflected $\theta$ row. They are joined by overlapping
the common
vertical bonds. (b) A happy bond (heavy line) has the $t$-values of the four
adjacent bonds equal in pairs, as shown.}
\end{figure}

At this point we have established a $T$ configuration for
which all the vertical bonds are happy. We have no
knowledge of the happiness of the zig-zag bonds between the
vertical bonds. Our next goal is to make $50\%$ of the
remaining zig-zag bonds happy without destroying the
happiness of very many of the vertical bonds. This is the
difficult step.

\emph{Step 2.}
We resist the temptation to cut the cylinder
along the line by which it was originally formed. Instead,
we proceed as follows, and we urge the reader to follow
this construction by forming a cylinder from a piece of
paper and tape. We start at one edge of the cylinder and,
with a pair of scissors we cut the paper along a line that
is $30^\circ$ from the axis (so $60^\circ$ from
the vertical direction). The line should not go through the
atoms and it should cut an even number of bonds. In this way
we cut a
spiral that eventually reaches the other edge of the
cylinder and we lay the paper out flat on the table. It
will be noted that there are two ways to tape this lozenge
piece of paper together to form a cylinder. One way is just
to put it back together in the way we cut it. The second
way, which is the one we will use, is to connect the
lozenge to itself with a shift in the $x$-direction. We pay
a surface energy to accomplish this. The old vertical bonds
that were parallel to the $y$-axis now lie at an
angle of $60^\circ$ to the axis of the new cylinder (which
we again assume to be parallel to the $x$-axis), and a
different set of bonds now lies parallel to the $y$-axis.

Again we focus our attention on a single row of the new cylinder, as shown in
Fig.~2(a). The happiness established in
Step 1 means that every second zig-zag bond of the new row
is happy. We now repeat the reflection positivity argument.
Inequality \eqref{eq:cb} allows us to repeat one row
periodically. This repetition preserves the happiness of
every second bond on the zig-zag line and makes every
vertical (in the new orientation) bond happy. The resulting
network of happy bonds consists of chains as shown by
heavy lines in Fig.~3(a).

\begin{figure}
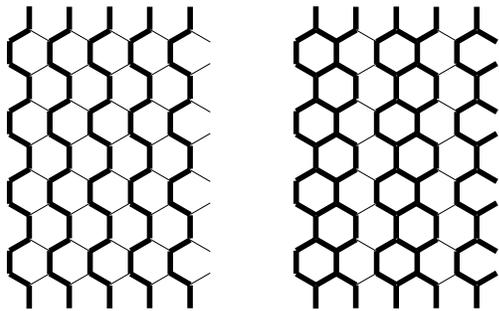

 \input{graph50.pstex_t}
\quad\quad\quad
 \input{graph75.pstex_t}
\caption{(a) The heavy lines in the left figure show the more than $50\%$
happy bonds after reflecting in orientations 1 and 2. (b) shows the more than
$75\%$ happy bonds after reflecting in orientations 1, 2 and again 1.}
\end{figure}

\emph{Step 3.}
We repeat Step 2 by cutting a spiral and retying the
cylinder as it was initially in Step 1. Since we have cut the chains
from Step 2 at an angle of $30^\circ$ relative to the
cylinder axis, we observe that every row parallel to the
cylinder axis has a zig-zag line with three out of every
four bonds happy. Reflection positivity \eqref{eq:cb}
allows us to repeat one such row endlessly. All the
resulting vertical bonds are happy and out of every
sequence of four zig-zag bonds three are happy. The
configuration of happy bonds again consists of chains but
now these chains are thicker; on every row the chain has
three zig-zag bonds instead of one. This is shown in Fig.~3(b).

\emph{Step 4.}
We cut and paste again and use reflection positivity to
achieve happiness for seven out of eight zig-zag bonds. The
chains are now seven bonds thick.

After $k$ steps all the vertical bonds and $2^k-1$ out of every
consecutive $2^k$ zig-zag bonds are happy. We have paid a surface energy
$k$ times and we also
might have incorrect bonds along the sequence of cut lines,
but each of these corrections can change the energy at most
by a quantity of order $L$. If we take $k\sim \ln_2 L$, we
find that all except possibly order $k L$ bonds are
happy, and we have paid an energy error of at most $O(L \ln_2 L)$.
This is what we wanted to show.

\emph{Conclusion}. The Hubbard model for graphene on the hexagonal lattice has
a special property called reflection positivity that the square lattice does
not have. With the aid of this property we are able to limit the kind of
lattice distortions allowed by this model. Only distortions that preserve the
three-fold reflection symmetry of the lattice are possible in the thermodynamic
limit.

We are indebted to G. Dunne, D. Abanin and A. Giuliani for illuminating
discussions and G. Holzegel for technical help. A grant from the
U.S.~National Science Foundation is acknowledged: PHY-0965859 (E.L.).


\bibliographystyle{amsalpha}

\end{document}